\documentclass{article}
\title{The ground state energy per site of the quantum and classical Edwards-Anderson spin glass in the thermodynamic limit}
\author{Walter F. Wreszinski\\
Instituto de F\'{\i}sica\\
Universidade de S\~{a}o Paulo\\
Caixa Postal 66318\\
05314-970 S\~{a}o Paulo, SP, Brasil}
\begin{document}

\maketitle
\begin{abstract}
We derive general rigorous lower bounds for the average ground state energy per site $e^{(d)}$ of the quantum and classical Edwards-Anderson spin-glass model in dimensions $d=2$ and $d=3$ in the thermodynamic limit. For the classical model they imply that $e^{(2)} \ge -3/2$ and $e^{(3)} \ge -2.204\cdots$.
\end{abstract}

\section{Introduction}
 
The study of ground states (GS) of short-range spin-glasses is a very difficult subject, which is demonstrated by the fact that finding the GS of a spin-glass in a three-dimensional lattice is an NP-complete problem \cite{Bar}: these systems are thereby placed among those with the highest computational complexity. After several exact results for finite lattices have been reported (see later), a fundamental problem poses itself : how far removed are such results from the thermodynamic limit? We propose to offer a first (partial) answer in this paper.

Our idea is to apply a method of Bader and Schilling \cite{BS} (for short BS method), which yields exact lower bounds to the GS energy per site for quantum Hamiltonians in the thermodynamic limit in terms of known results for finite lattices (cells), to random systems, such as the quantum anisotropic version of the Edwards-Anderson (EA) \cite{EA} spin-glass model treated in \cite{CL}, which includes the EA model as a special case. Our main observation is that the BS method is very naturally adapted to the concept of independence (of the couplings) in contrast to the standard methods of obtaining lower bounds, such as Temple's inequality \cite{T}, where an expectation value of the Hamiltonian appears in a denominator. Note that, in some instances, lower bounds given by the BS method may be exact: examples occur in the theory of quantum spin systems with competing interactions (\cite{BS},\cite{COW}), as well as in a theory of anisotropic ferromagnetic quantum domains \cite{ASW}, but in the present case this is not to be expected due to the aforementioned complexity. For two nice textbook accounts of rigorous methods in the theory of disordered systems, see \cite{Ne} and \cite{B}. 

Most of the numerical research on the EA spin glass ended at about 1980: we refer to Binder's review article \cite{Bi} for the results of Monte-Carlo simulations, which predict a GS energy per lattice site of about $-1.4$ for the two-dimensional model, and about $-1.9$ for the three-dimensional model, but, as far as we know, no exact results exist, except for finite lattices, which we shall refer to shortly. Interest remained, however, in connection to the possible applicability of Parisi's theory to the EA spin glass : see \cite{CGGPV} for a recent reference. Concerning lower bounds, Hoever and Zittarz \cite{HZ} report a surprising value of $-1.4219$, in a remarkable paper which, we believe, merits to be revisited. Although they refer to this bound as ''rigorous'', we believe that it is not, at this point, for a number of reasons: it would be so if the conjectured analyticity of their expansion held up to $T = 0$, but this is an open problem. Otherwise, it seems that a more careful estimation of the rest term after a given finite number of terms would be necessary, as well as a proof of the interchangeability of the limits $T \to 0$ and the thermodynamic limit in this expansion to finite order; this problem also occurs in connection with the ground state (or residual) entropy \cite{PP}. There are also important bounds due to Kirkpatrick \cite{K}, which will be commented upon in the conclusion.

\section{Results}

As in \cite{CL}, for each finite set of points $\Lambda \subset \mathbf{Z}^{d}$, where the dimension $d$ will be restricted to the values $d=2$ and $d=3$, consider the Hamiltonian
$$
H_{\Lambda}(\{J\}) = \sum_{i,j \in \Lambda} J_{i,j} \Phi_{i,j}
\eqno{(1-a)}
$$
where $\Phi_{i,j}$ with $i,j \in \Lambda$ are self-adjoint elements of the algebra generated by the set of spin operators, the Pauli matrices $\sigma_{i}^{x}$, $\sigma_{i}^{y}$, $\sigma_{i}^{z}$, $i \in \Lambda$, on the Hilbert space
${\cal H}_{\Lambda} = \otimes_{i \in \Lambda} \mathbf{C}_{i}^{2}$, given by
$$
\Phi_{i,j} = \alpha_{x} \sigma_{i}^{x}\sigma_{j}^{x} + \alpha_{y} \sigma_{i}^{y}\sigma_{j}^{y} 
+\alpha_{z}\sigma_{i}^{z}\sigma_{j}^{z}
\eqno{(1-b)}
$$
for $|i - j| = 1$, and zero otherwise. The random couplings $J_{i,j}$, with $|i - j| = 1$, are random variables (r.v.) on a probability space $(\Omega, {\cal F}, P)$, where ${\cal F}$ is a sigma algebra on $\Omega$ and $P$ is a probability measure on $(\Omega,{\cal F})$. We may take without loss of generality 
$$\Omega = \times_{B^{d}} S \eqno{(2-a)}$$ where $S$ is a Borel subset of $\mathbf{R}$, $B^{d}$ is the set of bonds in $d$ dimensions, and assume that the $J_{i,j}$ are independent, identically distributed r.v.. In this case, $P$ is the product measure $$dP = \times_{B^{d}} dP_{0} \eqno{(2-b)}$$ of the common distribution $P_{0}$ of the random variables, which will be denoted collectively by $J$. The corresponding expectation (integral with respect to $P$) will be denoted by the symbol $Av$. We have to assume that $$ Av(J_{i,j}) = 0 \eqno{(2-c)}$$ for all $i,j \in \mathbf{Z}^{d}$, i.e., that the couplings are centered. Let $E_{\Lambda}$ denote the GS energy of $H_{\Lambda}$, i.e., $E_{\Lambda} = \inf spec(H_{\Lambda})$. The following result was proved, among several others, in \cite{CL}:

\textbf{Theorem 1} (\cite{CL}) For $P$- a.e. $\{J\}$, the limit below exists and is independent of the b.c.:
$$
e^{(d)} \equiv \lim_{\Lambda \nearrow \mathbf{Z}^{d}} \frac{E_{\Lambda}}{|\Lambda|}
\eqno{(3-a)}
$$
and
$$
e^{(d)} = \inf_{\Lambda} \frac{E_{\Lambda}}{|\Lambda|}
\eqno{(3-b)}
$$
where $|\Lambda|$ denotes the number of sites in $\Lambda$. Finally,
$$
e^{(d)} \ge e^{(d+1)}
\eqno{(3-c)}
$$

(3-a) is the far-reaching property of self-averaging (see \cite{A} for a discussion) from which it follows that, $P$- a.e.,
$$
e^{(d)} = \lim_{\Lambda \nearrow \mathbf{Z}^{d}} \frac{Av(E_{\Lambda})}{|\Lambda|}
\eqno{(3-d)}
$$

Let $\Lambda_{N}$ denote a square with $N$ sites if $d=2$ or a cube with $N$ sites if $d = 3$, and write 
$H_{N}^{(d)}(J) \equiv H_{\Lambda_{N}}(J)$. We now adopt periodic b.c. for simplicity, but the final result is independent of the b.c. due to theorem 1. We may write
$$
H_{N}^{(d)}(J) = \sum_{n \in \Lambda_{N}} H_{n}^{(d)}(J)
\eqno{(4-a)}
$$
where $H_{n}^{(d)}$ is given by
$$
H_{n}^{(d)}(J) = c_{d} \sum_{(i,j) \in \Lambda_{n}^{(d)}} J_{i,j}\Phi_{i,j} 
\eqno{(4-b)}
$$
Above, $c_{d}$ are factors which eliminate the multiple counting of bonds, i.e.,
$$
c_{2} = 1/2
\eqno{(4-c)}
$$
and
$$
c_{3} = 1/4
\eqno{(4-d)}
$$
and $\Lambda_{n}^{(2)}$ is a square labelled by a site $n$, for which we adopt the convention, using a right-handed $(x,y)$ coordinate system, that $n=(n_{x},n_{y})$ is the vertex in the square with the smallest values of $n_{x}$ and $n_{y}$. Similarly, $\Lambda_{n}^{(3)}$ is a cube labelled by a site $n = (n_{x},n_{y},n_{z})$ with, by the same convention, the smallest values of $n_{x}$, $n_{y}$ and $n_{z}$. Due to the periodic b.c., the sum in (4-b) contains precisely $N$ lattice sites. The notation $H_{n}^{(d)}$ is short-hand for its tensor product with the identity at the complementary lattice sites in 
$\Lambda_{N}\backslash \Lambda_{n}^{(d)}$.
Let $E_{0}^{(N,d)}(J)$ denote the GS energy of $H_{N}^{(d)}(J)$ , $E_{0}^{(n,d)}(J)$ the GS energy of 
$H_{n}^{(d)}(J)$ and
$$
E_{0}^{(d)} \equiv Av(E_{0}^{(n,d)}(J))
\eqno{(4-e)}
$$
By the condition of identical distribution of the r.v. $J_{i,j}$, $E_{0}^{(d)}$ does not depend on $n$, which is implicit in the notation used. We have

\textbf{Theorem 2} The following lower bound holds:
$$
\frac{Av(E_{0}^{(N,d)}(J))}{N} \ge E_{0}^{(d)}
\eqno{(5-a)}
$$

\textbf{Proof}. By the variational principle,
\begin{eqnarray*}
E_{0}^{(N,d)}(J) = \inf_{(\psi,\psi)=1}(\psi, H_{N}^{(d)}(J) \psi) \ge\\
\ge \sum_{n \in \Lambda_{N}}\inf_{(\psi,\psi)=1}(\psi, H_{n}^{(d)}(J) \psi)
\end{eqnarray*}
$$\eqno{(5-b)}$$
Let $(\psi_{\alpha})$ denote the normalized eigenstates of $H_{n}^{(d)}(J)$ corresponding to eigenvalues 
$\{E_{\alpha}^{(n,d)}\}$ with $0 \le \alpha \le l_{n}$ and $ E_{0}^{(n,d)}(J) \le \cdots E_{l_{n}}^{(n,d)}(J) $ ,
 $l_{n}+1=2^{|\Lambda_{n}^{(d)}|}$ denoting the dimension of the Hilbert space over $\Lambda_{n}^{(d)}$. We may write the general normalized wave-function $\psi$ in 
${\cal H}_{\Lambda_{N}} = {\cal H}_{\Lambda_{n}^{(d)}} \otimes {\cal H}_{\Lambda_{N} \backslash \Lambda_{n}^{(d)}}$ in the form
$\psi = \sum_{\alpha,\beta} c_{\alpha,\beta}\psi_{\alpha}\otimes\phi_{\beta}$, with $\sum_{\alpha,\beta}|c_{\alpha,\beta}|^2=1$, and $(\phi_{\beta})$ an orthonormal basis of ${\cal H}_{\Lambda_{N}\backslash \Lambda_{n}^{(d)}}$. Since $H_{n}^{(d)}(J)$ equals the identity on the latter sites,
\begin{eqnarray*}
\inf_{(\psi,\psi)=1} (\psi, H_{n}^{(d)}(J) \psi) = \\
\inf_{c_{\alpha,\beta}}\sum_{\alpha,\beta}|c_{\alpha,\beta}|^2 E_{\alpha}^{(n,d)}(J)\\
\ge E_{0}^{(n,d)}(J)
\end{eqnarray*}
$$\eqno{(5-c)}$$
which, together with (5-b), yields
$$
E_{0}^{(N,d)}(J) \ge \sum_{n \in \Lambda_{N}} E_{0}^{(n,d)}(J)
\eqno{(5-d)}
$$
We now remark that , when performing the average over the $J$ on the r.h.s. of (5-d),  each r.v. 
$E_{0}^{(n,d)}(J)$ is the product of a function of the couplings in $\Lambda_{n}^{(d)}$ and the identity function of the couplings in $\Lambda_{N} \backslash \Lambda_{n}^{(d)}$; by mutual independence of the $J_{i,j}$ this average equals, for each $n$, an average restricted to the couplings in $\Lambda_{n}^{(d)}$ which equals the left hand side of (4-e): this proves (5-a). q.e.d.

\section{Application to the classical Edwards-Anderson spin glass}

The classical EA spin glass is the special case $\alpha_{x} = \alpha_{y} = 0$ in (1-b); we also set $\alpha_{z} = 1$. In this case $e^{(d)}$, given by the r.h.s. of (3-d), is invariant under the ''local gauge transformation'' 
$\sigma_{i}^{z} \to -\sigma_{i}^{z}$ together with $J_{i,j} \to -J_{i,j} \forall j | |j-i|=1$, whatever the lattice site $i$. Accordingly \cite{To} an elementary square (''plaquette'') $P$ is said to be frustrated (resp. non-frustrated) if 
$G_{P} \equiv \prod_{(i,j) \in P} J_{i,j} = -1$ resp. $+1$. Note that $G_{P}$ is gauge-invariant and that, for the quantum XY (or XZ) model defined by setting $\alpha_{y} = 0$ in (1-b), $e^{(d)}$ is also locally gauge-invariant if we add to the above definition the transformation $\sigma_{i}^{x} \to -\sigma_{i}^{x}$, i.e., the transformation on the spins is defined to be a rotation of $\pi$ around the $y$ - axis in spin space.

Since the Pauli z-matrices commute, finding the minimum eigenvalue of (1-a) in the classical EA case is equivalent to find the configuration of Ising spins $\sigma_{i} = \pm 1$, denoted collectively by $\sigma$, which minimizes the functional
$$
F(\sigma,J) \equiv \sum_{\sigma} J_{i,j} \sigma_{i} \sigma_{j}
\eqno{(1-c)}
$$ 
The minimal energy of a frustrated plaquette equals $E_{P,f} = -2$ and of a non-frustrated plaquette $E_{P,u} = -4$. We have, now:

\textbf{Proposition 1} Let $dP_{0}$ in (2-b) be the Bernoulli distribution $dP_{0} = 1/2(\delta_{J} + \delta_{-J})$ and set for simplicity $J = 1$. Then:
$$
e^{(2)} \ge -3/2
\eqno{(6-a)}
$$
and
$$
e^{(3)} \ge -1/4\frac{36096}{4096}
\eqno{(6-b)}
$$

\textbf{Proof}. For $d = 2$, there are exactly eight frustrated plaquettes (with $E_{P,f} = -2$) and eight non-frustrated plaquettes (with $E_{P,u} = -4$), hence by (4-b), (4-c) and (4-e),
$E_{0}^{(2)} = (1/2)(1/16)(-2\times8 - 4\times8) = -3/2$ and (6-a) follows from (5-a). For $d = 3$, there are $2^{12}$ possible configurations of the $J$ and $2^{8}$ possible values of the products $\sigma_{i}\sigma_{j}$, hence a total of $921600$ parameter values, almost one million, indeed huge in comparison with $256$ for $d = 2$. We obtain the r.h.s. of (6-b) by (4-b),(4-d) and the use of a convenient program which computes $E_{0}^{(3)}/c_{3}$, with $E_{0}^{(3)}$ given by (4-e). In order to do so, the program performs a finite number of operations involving just sums, subtractions and comparison of integers, hence the final result is exact. q.e.d.

\textbf{Remark 1} Due to (3-b), the approach to the thermodynamic limit is in a monotonically decreasing fashion. Thus any exact value for the ground-state energy per site for a finite sample is an upper bound to $e^{(d)}$, and, by (3-c), any upper bound to $e^{(2)}$ is an upper bound to $e^{(3)}$. It is, however, difficult to ascertain whether some of these ''exact values'' are rigorous. Those obtained by the so-called branch-and-bound algorithm are rigorous, because rigorous numerical error bounds of $O(1/L)$, for a $L \times L$ lattice, in two dimensions, may be derived: see ''Exact Ground States of Two-Dimensional $\pm J$ Spin Glass'', by C. De Simone, M. Diehl, M. J\"{u}nger, P. Mutzel, G. Reinelt, and G. Rinaldi, available at the web, in particular their Proposition 1.  A similar rigorous result \cite{HP} (we thank Prof. S. Homer for this reference) is also available at the web, yields the GS energy per site for a 5X5X5 lattice. Hence, by monotonicity and according  to the latter result,
$$
e^{(3)} \le -1.759
\eqno{(7-a)}
$$
and in view of the former,
$$
e^{(2)} \le -1.39\cdots
\eqno{(7-b)}
$$
where the number on the r.h.s. corresponds roughly to the best result estimated from their figure. An early reference to the case $d = 3$ is \cite{KK1}.  

\textbf{Remark 2} In \cite{D}, B. Derrida introduced the now famous random energy model (REM) as an approximation to the EA spin glass. In the case of the Bernoulli distribution, he obtains the lower bounds (see his eq. (54)):
$$
e^{(2)} \ge -1.560
\eqno{(7-c)}
$$
and 
$$
e^{(3)} \ge -1.956
\eqno{(7-d)}
$$
which are rigorous for the REM. It is interesting to remark that (7-c) compares well with (6-a) and (7-c) also compares well with (6-b), which may be written
$$
e^{(3)} \ge -2.204\cdots
\eqno{(6-b1)}
$$ 

\textbf{Remark 3} To the best of our knowledge, proposition 1 provides the first (nontrivial) rigorous lower bound both for $e^{(2)}$ and $e^{(3)}$. Using the natural misfit parameter $$m = \frac{|E^{id}| - |E_{0}|}{|E^{id}|} \eqno{(8)}$$ as a measure of plaquettes frustrated or bonds unsatisfied (see (4) of \cite{KK2}), where $E_{0}$ denotes the ground state energy of the frustrated system and $E^{id}$ is the ground state energy of a relevant unfrustrated reference system, we find from proposition 1 in the $d = 2$ case the lower bound $m \ge 0.25$ and for $d = 3$ the lower bound $m \ge 0.26\cdots$: thus, in both cases, the measure of frustrated plaquettes or unsatisfied bonds as defined above is at least of the order of $1/4$.

\section{Conclusion}

Consider the dual lattice, formed by the centers of the squares ($d = 2$) or cubes ($d = 3$) of the original lattice. Toulouse \cite{To} considered paths (strings) made up of lines joining neighboring points in the dual lattice and crossing the unsatisfied bonds, and suggested that the GS is characterized by the configuration(s) minimizing the average length $L$ of such  strings (in two dimensions, in three a much greater complexity arises and the average surface-to-perimeter ratio of ''minimal covering surfaces'' $\gamma$ (see \cite{K}) is the quantity to consider). Based on this suggestion, Kirkpatrick 
\cite{K}, in a beautiful paper, obtained the bounds (6-a) and
$$
e^{(3)} \ge -2.25
\eqno{(9)}
$$
under assumptions which seem to be eminently reasonable ($L \ge 1$ and $\gamma \ge 1/4$) but are actually very hard to prove. Since then, several suggestions that the emergence of certain global configurations (i.e., configurations  of the lattice as a whole), such as percolation of zero energy configurations (see \cite{SBLB}, \cite{BF}) or ''domain walls'' associated to low-energy excitations consisting of clusters of overturned spins (with respect to some ground state) \cite{FH}, with certain scaling characteristics, are responsible for the GS and low-temperature properties of spin glasses. We refer to 
\cite{BM} for a more complete list of references up to 1987. The fact that the r.h.s. of (6-b1) is even slightly better than (9), which relied on heuristics based on a global property, suggests that \textbf{local} properties might play an equally important role in the low-temperature physics of short-range spin glasses. It must be emphasized that for $d = 3$ there exist local correlations between the frustrated plaquettes, for instance, for $d=3$, the product of the $G_{P}$ over all plaquettes in the unit ''cell'' $\Lambda_{n}^{(3)}$ equals $+1$, therefore there must be an even number of frustrated (and unfrustrated) plaquettes in a cell configuration. Therefore, unlike the case $d=2$, not every plaquette configuration occurs independently in one cell configuration.

We finish this paper with two remarks. In a low temperature expansion, the first term in the partition function is the GS degeneracy. Since Parisi's seminal work \cite{MPV}, it is assumed that this degeneracy becomes infinite in the thermodynamic limit, and, indeed, it was recently shown numerically \cite{CGGPV} that for the EA spin-glass with Bernoulli random couplings the overlap introduced by Parisi \cite{MPV} is a good order parameter. In this connection, there exist rigorous lower bounds for the entropy density at zero temperature for $d = 2$ and $d=3$: the first was due to \cite{ARS}, which was greatly improved by \cite{PP}. For $d = 3$, see \cite{F}.

The second remark concerns the quantum model (1), to which the method is applicable: for $\alpha_{y} = 0$ we have the (quantum) XY model, which may define a different universality class, and it would therefore be very interesting to see how the present bounds would change with $\alpha_{x}$ (fixing $\alpha_{z} = 1$). It is also, conceptually as well as practically, important to view the Ising model (1-c) as an anisotropic limit (e.g., as $\alpha_{x} \to 0$ and $\alpha_{y} \to 0$ in (1-b)) of quantum models: indeed, the reason why the critical exponents of the Ising model in three dimensions (see, e.g., \cite{JZ})  are so close to those measured in real magnetic systems is that most of the latter are highly anisotropic. Thus we feel that it is important that general proofs are stable by small quantum perturbations. In addition,  the only natural (i.e., not imposed) dynamical evolution is the quantum evolution. One exact example of the latter is furnished by the model treated in \cite{W}, to which we also refer for a discussion (with references) of the role of probability distributions in random systems.

\textbf{Acknowledgements} We are very grateful to Jo\~{a}o L. M. Assirati for the ingenious program which allowed to compute the r.h.s. of (10a), to Pierluigi Contucci for an enlightening correspondence on the result of \cite{CL} and for referring us to ref. \cite{PP}, and to the referees for helpful and constructive remarks.

\end{document}